\documentclass{article}

\usepackage[english]{babel}

\usepackage[letterpaper,top=2cm,bottom=2cm,left=3cm,right=3cm,marginparwidth=1.75cm]{geometry}
\usepackage[symbol]{footmisc}
\usepackage{tocbibind}
\usepackage[toc,page]{appendix}
\usepackage{subfigure} 
\usepackage{authblk}

\usepackage{amsmath}
\usepackage{graphicx}
\usepackage[colorlinks=true, allcolors=blue]{hyperref}

\title{Opinion formation by belief propagation: A heuristic to identify low-credible sources of information}
\author{Enrico Maria Fenoaltea$^{1}$\footnote{Corresponding author. Email: enricofenoaltea@hotmail.it} and Alejandro Lage-Castellanos$^2$}
\date{}
\begin{document}
\maketitle

%
%
%
\begin{center}
$^1$Physics Department, University of Fribourg, Chemin du Mus\'ee 3, 1700 Fribourg, Switzerland\\
$^2$Group of Complex Systems and Statistical Physics, Physics Faculty, Havana University, La Habana, CP 10400, Cuba\\
\end{center}

\begin{abstract}
With social media, the flow of uncertified information is constantly increasing, with the risk that more people will trust low-credible information sources. To design effective strategies against this phenomenon, it is of paramount importance to understand how people end up believing one source rather than another. To this end, we propose a realistic and cognitively affordable heuristic mechanism for opinion formation inspired by the well-known belief propagation algorithm. In our model, an individual observing a network of information sources must infer which of them are reliable and which are not. We study how the individual's ability to identify credible sources, and hence to form correct opinions, is affected by the noise in the system, intended as the amount of disorder in the relationships between the information sources in the network. We find numerically and analytically that there is a critical noise level above which it is impossible for the individual to detect the nature of the sources. Moreover, by comparing our opinion formation model with existing ones in the literature, we show under what conditions people's opinions can be reliable. Overall, our findings imply that the increasing complexity of the information environment is a catalyst for misinformation channels.
\end{abstract}

\textbf{keywords:} Belief propagation, Signed network, learning, Heuristic, node classification, Phase transition

\section{Introduction}

Social media have revolutionized how people communicate and inform themselves, becoming the primary source of information for most users. However, content in social media is often published without the intermediation of experts, thus increasing the risk of spreading unreliable news \cite{lazer2018science, shao2018spread, guess2020exposure, gallotti2020assessing}.
With the advance of the research on how opinions propagate in networks \cite{del2016spreading, hu2018local}, how news goes viral \cite{bessi2015science, juul2021comparing}, and how polarization affects public opinion \cite{schmidt2017anatomy, johnson2020online, vicario2019polarization}, methods have been proposed to limit the spread of incorrect information and help users distinguish true news from fake news \cite{nasir2021fake, zollo2017debunking, gupta2014tweetcred, ratkiewicz2011detecting}. Despite these efforts, misinformation still threatens society. Indeed, it is not trivial for the general public to distinguish between reliable and unreliable content. A pertinent example was the recent media circus around the war in Ukraine: people have faced an overwhelming amount of both reliable news and news whose only purpose was to make propaganda. This has generated widespread confusion with dangerous implications for society. Therefore, it is crucial to understand how people form their opinions and provide them with strategies for properly getting informed in a world full of often conflicting information channels.

In addition to studies in psychology \cite{leeper2014political, gerard1987dynamics} and sociology \cite{friedkin2017truth, friedkin1990social}, opinion formation is widely investigated by quantitative methods, such as network science \cite{meng2018opinion, jalili2013social}, game theory \cite{acemoglu2011opinion, di2007opinion}, or statistical physics \cite{baumann2020modeling, pham2021balance, schweitzer2018sociophysics}. In particular, models have been proposed to study opinion contagion in social networks \cite{alvarez2016network, juul2019hipsters}, and how such contagion leads to the emergence of cohesion \cite{redner2019reality, fenoaltea2023phase} or fragmentation \cite{pham2022empirical, meng2023disagreement} in society. Recently, a model has been proposed in \cite{medo2021fragility} that differs from the models mentioned above by including potential relationships (both positive and negative) between the subjects on which an individual, in the absence of social influence, tries to form an opinion. In fact, such connections are a fundamental component of the world of social media and information channels \cite{easley2012networks, hansen2010analyzing}. The authors in \cite{medo2021fragility} discuss a cognitively and computationally simple heuristic rule by which an individual navigates the network of subjects and gradually forms an opinion about each of them. Then, based on an a priori hidden truth, the proportion of correct and incorrect opinions is measured. Other heuristic rules were then investigated in \cite{meng2021whom}. Such works do not consider the probabilistic and uncertain nature of human decisions \cite{edwards1962dynamic, pouget2016confidence}, but assume that an individual's opinions have a binary character (e.g., trust or distrust of an information channel). However, people's beliefs and opinion systems can be more nuanced.

Within the same framework of these models (which we explain in more detail below), here we allow for uncertainties in the opinion formation process by studying a new heuristic rule whose mechanism is based on the well-known belief propagation algorithm \cite{pearl1988probabilistic}. In particular, we consider an individual who wants to learn something about a topic (e.g., the motivations for the Ukrainian war)  and observes a network of information sources- some reliable, others unreliable- discussing that topic. While it is immediate to understand that two sources promote conflicting content, it is more difficult for the layman to choose which one to trust. These sources of information can be websites, Youtube channels, or newspapers whose interconnections represent their mutual relationship of trust or distrust. For example, two Youtube channels with the same view on a subject will tend to collaborate; conversely, if they have an opposite opinion, they will diss each other. In the literature, such a relationship system is formalized as a signed network with positive and negative links \cite{harrigan2020negative}. Moreover, the relationships between sources of information should follow the logic of Heilder's balance rule \cite{heider1946attitudes}: if two sources have a link of the same sign with a third source, they are connected to each other by a positive link; conversely, if they have links of the opposite sign with a third source, they are connected by a negative link (this is a formalization of the proverb ``my enemy's friend is my friend"). While this is certainly an oversimplification of the more complex existing spectrum of relationships between different sources of information, this structure has been observed in data sets of various political and social systems \cite{moore1979structural, fontan2021signed, facchetti2011computing}. Nevertheless, there may be several deviations: for example, for some reason, two sources of information that generally agree may disagree on a minor issue, and the individual is misled because he/she observes a negative connection between them; or the topic addressed is so complex that it is difficult to determine which is the true relationship between the sources, increasing the risk that the individual will observe incorrect link signs. To account for this, we assume that the link signs in the network of information sources are at odds with the balance rule with some small probability that we refer to as noise.

Assuming that, initially, the individual has information about the reliability of only one source of information in the network, his/her goal is to identify which of the sources are reliable by observing the (noisy) relationships between them. This task can be cognitively very expensive, so humans must rely on simple heuristics when dealing with complex issues \cite{daniel2017thinking, gigerenzer2011heuristic}. Hence, we study a probabilistic and local heuristic rule (i.e., where only local knowledge of the source network is required) inspired by the belief propagation algorithm. This algorithm is useful to efficiently solve inference problems by passing local messages \cite{yedidia2003understanding}. It has important applications in statistical physics \cite{yedidia2000generalized, altarelli2014bayesian}, error-correcting coding theory \cite{kschischang2001factor}, computer vision \cite{frey1998graphical}, and artificial intelligence \cite{mceliece1998turbo,pearl1988probabilistic}. In this paper, we recast the belief propagation algorithm in the framework of opinion formation. 

Specifically, we investigate how the noise affects the performance- in terms of the number of information sources correctly labeled as reliable or unreliable- of our belief propagation rule and show that, in addition to realistically describing how humans reason, it outperforms existing heuristics in the literature. However, we find analytically that there is a critical noise level beyond which, even with our belief propagation approach, one cannot extract information from the source network.
Then we show numerically that, independently of the noise, the performance of any local rules inevitably decreases as the size of the sources network increases. This adverse outcome can be limited if the individual forms his/her opinions by aggregating information from multiple sources. We also extend our model and the models in \cite{medo2021fragility, meng2021whom} to study the effect of repeated opinion updating, showing that it is beneficial only for local heuristic rules involving the above-mentioned multiple sources aggregation. 

 Our findings suggest that, because we have limited information capabilities, our opinions can be highly sensitive to the complexity of the information environment, thus calling for a policy to provide less cluttered communication to the general public. Moreover, by giving insights into what conditions increase the risk of people relying on low-credible sources, our work can help design new tools or improve existing ones to prevent the proliferation of false beliefs.

\section{Opinion formation processes}
\subsection{The basic framework}
Consider a system composed of $N$ interconnected sources of information. For simplicity, we assume that each source can be of two types, reliable or unreliable, with equal probability. We represent these two types with positive and negative signs, respectively. If two information sources of the same sign are connected, they have a positive relationship (mutual trust); if they are of the opposite sign, they have a negative relationship (mutual distrust). In this way, the information sources define an undirected signed network with $N$ nodes $V=\{i, i \in [1,N]\}$. Each node $i$ can be positive or negative, i.e., $s_i \in \{-1,1\}$, and the relationships between nodes are defined by a weighted graph where, whenever there is an edge between two nodes $i$ and $j$, this edge has an associate weight $J_{ij}=s_is_j$. This can be equivalently encoded into a weighted matrix $J$, whose elements $J_{ij}$ are zero if the sources are non-interacting, while +1 if they are of the same type and -1 if they are of a different type.
In the following, we consider only the random network topology with average degree $k \in \mathbf{N}$, which means that two nodes have a link with probability $p=k/(N-1)$. Therefore, the average number of links $E$ is given by $E=pN(N-1)$. We will refer to this network as the \textit{ground truth network} $G(V,E,J)$.

Now, consider an external observer who, having only partial knowledge of the nodes in $G$, tries to figure out which sources of information are reliable and which are not, i.e., forms opinions about them. Formally, The opinion $o_i$ on a node $i$ is a two-dimensional vector such that:
\begin{equation}\label{eq:opinion}
    o_i := \left[\begin{array}{c}
     P(s_i=+1) \\
     P(s_i=-1) 
\end{array} \right]
\end{equation}
where the first element represents the degree of confidence (or the probability), according to the external observer, that the information source $i$ is reliable, while the second element is its complement, i.e. $P(s_i=-1)=1-P(s_i=+1)$. We assume that the observer initially knows only one node $i^*$, such that $o_{i^*}=(1,0)$ if $s_{i^*}=+1$ and $o_{i^*}=(0,1)$ if $s_{i^*}=-1$. We refer to this node as \textit{seed node}.  

Then, we introduce a parameter $r \in [0,1/2]$ that represents the noise on the relation matrix $J$. Specifically, the observer is not able to "see" the real $J$ but observes a relation matrix $J^o$ such that
\begin{equation} \label{eq:probJij}
    \forall_{(i,j)} \quad J^o_{ij} = \left\{\begin{array}{cc}
        J_{ij} & \mbox{ with prob } 1-r  \\
        -J_{ij} & \mbox{ with prob }  r
    \end{array}\right.
\end{equation}
In this way, $r$ serves to include the structural noise present in any real system in which it is not straightforward to determine the type of relationship between two sources. When $r=0$, the matrix $J$ and the observed matrix $J^o$ coincide, so the observer knows with certainty which are the true relationships between the information sources. When $r=1/2$, however, $J^o$ is not informative about the links' signs in $G$, as $J$ and $J^o$ are not correlated.

Given a ground truth network $G(V,E,J)$, a seed node $i^*$, and a noise level $r$ generating $J^o$, the observer's goal is exploiting the information from $i^*$ and $J^o$ to form opinions that reflect as accurately as possible the true types of the nodes in $G$ (see Fig.\ref{fig:network}a). At the two extremes, the formation of an accurate opinion is either impossible ($J^o$ uninformative with $r=1/2$) or trivial ($J^o=J$ with $r=0$). Indeed, starting from the seed node, the observer can assign each node the type according to $\Pi_{i}J_{\pi_{i-1}\pi_i}$, where $\pi$ is any path connecting the seed node to the target node, and $\pi_i$ is the $i$-th node of the path $\pi$.

In the following, we introduce different opinion formation processes and study their performance as $r$ varies. As a measure of process performance, we use the resulting average overlap $q$ between the expected value (from the observer's point of view) of the sign of node $i$ and its true value $s_i$: 
\begin{equation} \label{eq:overlap}
    q := \frac{1}{N}\sum_{i=1}^N \langle o_i\rangle s_i,
\end{equation}
where $\langle o_i\rangle:= P(s_i=+1)-P(s_i=-1)$. The overlap is one when the observer infers with certainty the correct sign of each node of $G$ and is zero in the extreme case where the observer fails to extract any information about $G$.

\subsection{The optimal Bayesian}
Before introducing the opinion formation processes of an external observer able to handle only local information, we briefly describe the Bayesian approach, already examined in \cite{meng2021whom}. This will be useful as a conceptual introduction to the belief propagation algorithm.

Let us denote by $\{\textbf{s}\}$ the set of all possible configurations of node signs in $G$, and by $\textbf{s}$ a specific configuration. The conditional probability $P(J^o|\textbf{s})$ that, given a configuration $\textbf{s}$, a specific $J^o$ is observed, is
\begin{equation}
 P(J^o|\textbf{s})=\prod_{(i,j)\in E}\Psi(J^o_{ij}|J_{ij}),    
\end{equation}
where the product is over all the connected nodes in $G$ and $\Psi$ is the probability of observing $J^o_{ij}$ given $J_{ij}$. From Eq.~\ref{eq:probJij}, $\Psi(J^o_{ij}|J_{ij})=1-r$ if $J^o_{ij}=J_{ij}$, and $\Psi(J^o_{ij}|J_{ij})=r$ otherwise. The Bayes rule gives the conditional probability $P(\textbf{s}|J^o)$ that, given observations $J^o$, the true configuration of the nodes' signs in $G$ is $\textbf{s}$:
\begin{equation}\label{eq:bayes}
P(\textbf{s}|J^o)=\frac{P(J^o|\textbf{s}) P(\textbf{s})}{\sum_{\textbf{s}\in \{\textbf{s}\}}P(J^o|\textbf{s}) P(\textbf{s})}, \end{equation}
where $P(\textbf{s})$ is the \textit{a priori} distribution of the nodes' signs in $G$, and the normalization is given by the sum over all possible configurations.

From Eq.~\ref{eq:opinion}, the opinion $o_i$ about node $i$ is obtained from the probability $P(s_i=+1)$ that node $i$ is positive. Denoting by $\{\textbf{s}^i\}$ the set of all configurations in which $s_i=+1$, we can write:
\begin{equation}
  P(s_i=+1)=\sum_{\textbf{s}\in \{\textbf{s}^i\}}P(\textbf{s}|J^o).
\end{equation}
With this probability, we can also compute the average overlap as in Eq.~\ref{eq:overlap}.

The Bayesian approach described in this section is the optimal way to process available information \cite{meng2021whom}. Nevertheless, its computation is hard and its application is unfeasible from a cognitive point of view. Indeed, it requires the knowledge of all the configurations in $\{\textbf{s}\}$: since each node can be in two states (positive or negative), the number of configurations is $2^N$. When $N$ is large enough, this is a computationally prohibitive number even for computers.

In the following, we propose a heuristic opinion formation process where, contrary to the Bayesian approach, the amount of information to be processed is local and reasonable for a human observer.

\subsection{Belief Propagation and opinion formation} \label{sec:BP}
\begin{figure}[h!]
    \centering
   \subfigure[]{\includegraphics[width=0.45\linewidth]{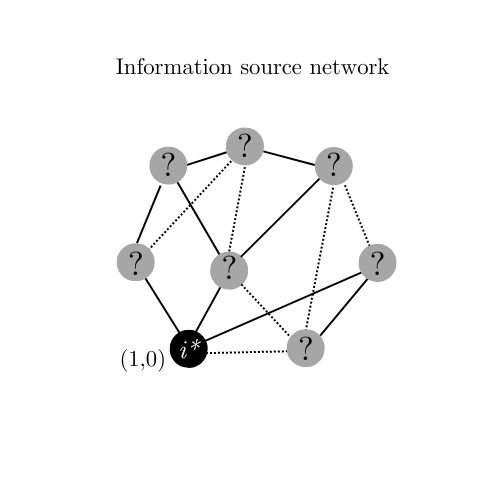}}
   \subfigure[]{\includegraphics[width=0.45\linewidth]{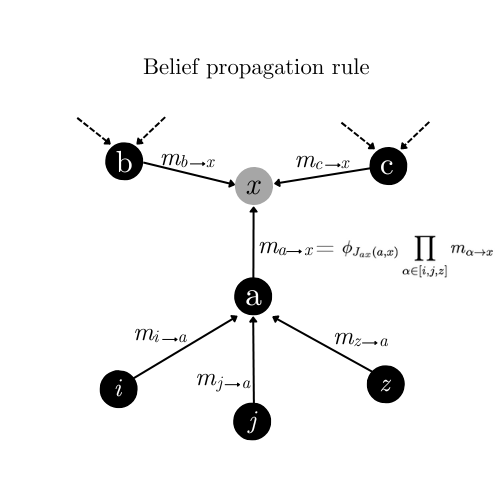}}
    \caption{(a) An instance of the information source network: the black node represents the seed node $i^*$ on which the observer has the opinion $o_{i^*}=(1,0)$ (in the figure, we assume that $s_{i^*}=+1$); the gray nodes are the sources on which the observer will have to form an opinion; the solid lines represent the positive links between sources while the dashed lines represent the negative links between sources. Note that the figure depicts the relationships observed by the individual (not those in the ground truth network). Indeed, there are some triangles that violate Heilder's balance rule. (b) A graphical representation of the message passing rule in the belief propagation algorithm: the message from node $a$ to node $x$ is given by the product of the messages toward $a$, excluding the one from $x$, multiplied by the compatibility matrix in Eq.\ref{eq:phi_matrix}. Finally, the opinion on $x$, $o_{x}$, is given by the product of all messages toward $x$.}\label{fig:network}
\end{figure}

We now introduce an opinion formation process, based on the well-known belief propagation (BP) algorithm \cite{yedidia2003understanding}, where the observer walks through the network starting from $i^*$ and gradually forms an opinion on each of the remaining nodes $i\ne i^*$. The BP algorithm is based on the so-called \textit{message passing}. We first introduce the message $m_{i \to j}$ from node $i$ to node $j$ as a two-component vector 
\[m_{i\to j} = \left(\begin{array}{c}
     m_{i \to j}^+ \\
     m_{i \to j}^- 
\end{array} \right), \]
where the first element can intuitively be understood as the observer's confidence stemming from the information about node $i$ that node $j$ is positive; similarly, the second element indicates the observer's confidence that node $j$ is negative. The message from node $i$ to node $j$ is defined recursively as the aggregation of all messages toward $i$ without the one coming from $j$ (see Fig.\ref{fig:network}b):
\begin{equation}\label{eq:message}
    m_{i\to j}=\phi_{J_{ij}}(i,j) \prod_{z\in\partial i \setminus j}m_{z \to i}.
\end{equation}
Here, $\partial i\setminus j $ is the set of nodes, excluding $j$, connected to $i$ in $G$. Notice that the product of the messages is a Hadamard product \cite{styan1973hadamard}, i.e., a component-wise product. $\phi_{J_{ij}}(i,j)$ is called the compatibility matrix between node $i$ and node $j$ and is defined as
\begin{equation*}
\phi_{J_{ij}}(i,j)=
\begin{pmatrix}
1-r & r\\
r &  1-r
\end{pmatrix} \quad \text{if $J^o_{ij}=+1$}
\end{equation*}
\begin{equation}\label{eq:phi_matrix}
\phi_{J_{ij}}(i,j)=
\begin{pmatrix}
r & 1-r\\
1-r &  r
\end{pmatrix} \quad \text{if $J^o_{ij}=-1$}
\end{equation}
and the product between this matrix and the vectors in Eq.\ref{eq:message} is the standard matrix product.

The compatibility matrix must be interpreted as the Bayesian information about the sign of the link between node $i$ and node $j$ and is the local analogue of Eq.\ref{eq:bayes}. Specifically, the terms in the diagonal of $\phi(i,j)$ represent the probability that nodes $i$ and $j$ have the same sign given the observed value of $J^o_{ij}$; Similarly, the terms in the anti-diagonal represent the conditional probability that $i$ and $j$ have opposite signs. 

Finally, the observer forms its opinion $o_i$ on node $i$ by multiplying all its incoming messages:
\begin{equation}\label{eq:opinion_comp}
 o_i= \frac{\prod_{z \in \partial i}m_{z\to i}}{Z},
\end{equation}
where $\partial i$ is the complete set of nodes connected to $i$ in $G$, and $Z$ is the normalization given by
\begin{equation}\label{eq:nor_message}
    Z:=\left(\prod_{z \in \partial i}m_{z\to i}^+\right)+\left(\prod_{z \in \partial i}m_{z\to i}^-\right).
\end{equation}

Now, in the line of thought of \cite{medo2021fragility}, we can define our BP process of opinion formation. Since at the beginning of the process the observer knows no nodes except the seed node $i^*$, at time $t=0$ we set $m_{i\to j}=(1/2,1/2)$ for any $i,j\ne i^*$, and $m_{i\to i^*}=o_{i^*}$ for any $i$. The process continues as follows:
\begin{enumerate}
    \item at each time step $t$, choose a random node $i$ with $o_i=(1/2,1/2)$ that has at least one neighbor on which the observer has an opinion different than $(1/2,1/2)$;
    \item compute the messages toward $i$, i.e. $m_{j\to i}$ for any $j\in \partial i$, as in Eq.~\ref{eq:message};
    \item the observer's opinion $o_i$ on $i$ is computed as in Eq.~\ref{eq:opinion_comp};
    \item return to step 1.
\end{enumerate}

This process is repeated until $t=N-1$, when the observer has an opinion on each node. In reality, however, people occasionally re-evaluate their beliefs. We can include this re-evaluation in the model by running the opinion formation process even for $t \ge N$ so that the observer evaluates more times the same node. For $t \ge N$, the first step of the process is modified as follows:
\begin{enumerate}
    \item at each time step $t \ge N$, choose a random node $i\ne i^*$.
\end{enumerate}
This process is repeated until a desired time $t$ is reached.

In this way, opinion formation undergoes two different phases: when $t<N$, the observer covers the entire information source network by exploring only the unknown nodes; when $t\ge N$, the observer can jump randomly on any nodes, as all nodes have already been explored in the past. When the observer passes over the node $i$ several times, it means that he/she is updating his/her opinion about $i$. Such an updated opinion is based on more information than the opinion formed for the first time, as in the former case the messages coming from all $i$'s neighbors have already been computed in previous steps. The value $\tau=t/(N-1)$ thus represents the average number of times an opinion is updated, and we shall refer to it as the observer's \textit{thinking time}.
We want to study the average overlap $q$ as a function of noise $r$, network size $N$, and thinking time $\tau$. A related model, involving the classification of nodes in a network into distinct groups based on network topology observation, is known in the literature as the stochastic block model. Its equilibrium state, corresponding to the infinite time case, has been extensively studied using the cavity method from statistical mechanics \cite{decelle2011inference,decelle2011asymptotic}. It is worth mentioning that, for finite-size networks, the cavity method is equivalent to belief propagation. Nevertheless, although our approach bears similarities to the stochastic block model, it incorporates the relationship between groups through the sign of the links, whereas the stochastic block model considers the different probability of link presence between elements of different groups: For an individual observer browsing a network of information sources, they are a very different problem. 

It is now worth dwelling on the interpretation of our BP opinion formation mechanism. In fact, at first glance, one might think that the computations described in this section are too intricate to be performed in our everyday reasoning. Here are some remarks to explain the connection between the belief propagation algorithm and human reasoning. First, let us elucidate the role of the compatibility matrix in Eq.\ref{eq:phi_matrix}. When we observe the current relationship between two sources of information, we are aware, based on our experience, that it gives insights into the true relationship between the two parties but, at the same time, also that such insights can be misleading.
For example, consider the case of anti-vax people: how likely is it that two individuals, both anti-vax (thus with $J^o_{ij}=+1$), also have the same general worldview (i.e., $J_{ij}=+1$)? The compatibility matrix answers this question by incorporating the observer's awareness of the reliability of the observed relations.
In this case, it is well known that supporting conspiracy theories about vaccines is strongly correlated with supporting populist political beliefs and distrusting of elites and experts \cite{hornsey2018psychological, motta2018knowing, kennedy2019populist}.
Naturally, this is not always the case: assuming that the noise is $r=0.1$, this is true $90\%$ of the time. In Bayesian reasoning, this data represents the likelyhood of $J_{ij}=+1$ or $J_{ij}=-1$ given a fixed $J^o_{ij}$. In turn, this is used by the observer to infer the posterior probability that $J_{ij}=+1$ or $J_{ij}=-1$ given the observed $J^o_{ij}$, corresponding to the elements of $\phi_{J_{ij}}(i,j)$. 

Once the posterior distribution of the link's sign connecting a node $i$ to a node $j$ is known, Eq.\ref{eq:message} corresponds to aggregating all the information on node $i$ from its neighbors. Thus, by combining the information on $i$ with the probabilities in $\phi_{J_{ij}}(i,j)$, the observer can extract information about $j$, with the support of the balance rule postulate (i.e., two nodes have a positive relationship if they have the same sign and a negative relationship otherwise). The final opinion on a node is then the aggregation of all messages directed to that node.

Unlike in the optimal Bayesian approach, in the BP approach, the observer needs only local information. Indeed, to form an opinion on a node, the observer must only know its neighbors. Hence, our BP heuristic is a feasible strategy for a human observer with limited computation capabilities. Moreover, there is strong behavioral and physiological evidence that the brain represents probability distributions and performs local Bayesian inference \cite{sanders2016signatures, sanders2016signatures, gershman2014amortized, pouget2013probabilistic}. Therefore, it is a plausible hypothesis that our BP mechanism may occur even spontaneously during individuals' opinion formation process. 

\subsection{Other heuristic processes}
As a benchmark, we also study and extend other processes proposed in the literature. In particular, we focus on the so-called \textit{majority rule} (MR) and \textit{random neighbor} (RN) processes \cite{medo2021fragility,meng2021whom}. 

\subsubsection*{Majority rule process}
With the basic framework described above, the MR process is defined as follows \cite{meng2021whom}. At time $t=0$, the opinion on the seed node is $o_{i^*}$; while for each $i\ne i^*$, the initial opinion is $o_i=(0,0)$. The following steps are iterated:

\begin{enumerate}
    \item at each time step $t$, choose a random node $i$ with $o_i=(0,0)$ that has at least one neighbor on which the observer has an opinion different than $(0,0)$;
    \item for any $j \in \partial i$ such that $o_j\ne (0,0)$, compute its vote $v_j$:
     \begin{equation*}
  v_j=\begin{cases} 
    +J_{ij}^o & \mbox{if }o_j=(1,0)
        \\
        -J_{ij}^o & \mbox{if }o_j=(0,1)
  \end{cases}
 \end{equation*}
    \item if $\sum_{j}v_j>0$, then $o_i=(1,0)$. If $\sum_{j}v_j<0$, then $o_i=(0,1)$. If $\sum_{j}v_j=0$, then $o_i=(1,0)$ or $o_i=(0,1)$ with equal probability;
    \item return to step 1.
\end{enumerate}

These steps are repeated until all opinions are different from $(0,0)$, that is, until $t=N-1$. 

\subsubsection*{Random Neighbor process}
At time $t=0$, the RN process has the same initial settings as the MR process. Then, it proceeds as follows \cite{medo2021fragility}:
\begin{enumerate}
    \item at each time step $t$, choose a random node $i$ with $o_i=(0,0)$ that has at least one neighbor on which the observer has an opinion different than $(0,0)$;
    \item chose a random node $j \in \partial i$ such that $o_j\ne (0,0)$.
    \item if $J^o_{ij}=+1$, then set $o_i=o_j$. If $J^o_{ij}=-1$, then set $o_i=(1,1)-o_j$;
    \item return to step 1.
\end{enumerate}
Again, these steps are repeated until $t=N-1$. 

We introduce the thinking time $\tau$ also for MR and RN processes. When $\tau>1$, the first step is modified for both as follows:
\begin{enumerate}
     \item at each time step $\tau \ge 1$, choose a random node $i\ne i^*$.
\end{enumerate}
The next steps, instead, are the same as before and processes continue until a desired $\tau$.

\subsection{Interpretations and differences between the proposed opinion formation processes} \label{sec:dif}

To elucidate the structural differences between the three heuristics presented in this paper, some remarks are in order.
First, in the RN process, the observer forms his/her opinions on a node $i$ by using the information from only one of its neighbors, whereas in both the MR and BP process the observer aggregates the information from all $i$'s neighbors. Hence, for the RN process, we expect a lower overlap $q$ (given the same amount of noise). On the other hand, the observer with an RN strategy demands the least cognitive and computational effort, while the cost of the BP and MR strategies grows with the network's average degree $\langle k \rangle$. Then, the difference between the aggregate information of the MR and that of the BP processes lies in the weight assigned to the information from each $i$'s neighbor. Specifically, in the MR process, the observer gives the same weighs to all $i$'s neighbors' votes; in our BP process, instead, a message from a $i$'s neighbor with $|\langle o\rangle|$ close to one (i.e., with small uncertainty), is more informative and thus has a larger weight than a message from a $i$'s neighbor with $|\langle o\rangle|$ close to zero (i.e., with higher uncertainty). Therefore, the BP process should outperform the MR process.

Second, in the MR and RN process, the opinion on a node can only be in two states: either $(1,0)$ or $(0,1)$. So, in this case, the overlap in Eq.\ref{eq:overlap} is simply the difference between the average number of nodes where the observer’s opinion matches the ground truth and that where it is opposite. Instead, in the BP process, the opinion is probabilistic (see Eq.~\ref{eq:opinion}) and admits a wider spectrum of possibilities. However, the interpretation of the overlap $q$ is the same as in the case of the MR or RN mechanism provided that, at the end of each realization of the process, the observer is forced to fix the value of each node using the marginal values as the decision criteria. In practice, for any $i$, the observer chooses $o_i=(1,0)$ with probability $P(s_i=+1)$, and $o_i=(0,1)$ otherwise.  
 
Last but not least, in both the MR and RN cases, opinion formation occurs with the observer unaware of the noise in the system. As we discussed in section \ref{sec:BP}, instead, the observer in the BP process exploits the knowledge of the noise to compute the posterior distributions in the compatibility matrix $\phi_{J_{ij}}(i,j)$. To obtain this knowledge, the observer has to undertake more extensive searches, and it is reasonable to assume that this involves more cognitive effort.

In summary, all three processes proposed in this paper require having only local information while exploring the network of information sources. However, the RN process is the least expensive but also the least accurate of the three; in contrast, the BP process requires the most effort but we expect it to pay off with the best performance; the MR process lies in between.  

\section{Results and discussion}
To perform our analysis, it is necessary to consider the \textit{a priori} distribution $P(\textbf{s})$ of the nodes' signs in $G$. Since in each realization of the source network, $s_i$ is +1 or -1 with equal probability, we have $P(\textbf{s})=1/2^N$. This induces the wrong idea that we are facing a highly complex model (NP-complete optimization problem, or spin glass like \cite{mezard2001bethe, mezard1987spin, marinari1998numerical}). However, in our analytical calculations, we use the fact that it is computationally and statistically equivalent to a simpler model that, using the physics terminology, belongs to the category of ferromagnetic models \cite{peierls1936ising}\footnote{The relationship between the noise $r$ and the temperature $T=1/\beta$ is given by $r(\beta)=\frac{e^{-\beta}}{2cosh(\beta)}$. It is a particularity of the current model that the parameter $r$ defines both the disorder of the system (through Eq.\ref{eq:probJij}) and its temperature. Disordered models can be hard to treat, especially if the disorder is high and the temperature is low. However, in our case, $r=0$ produces a $T=1/\beta = 0$ temperature but in a fully ferromagnetic model, which is easy because it has no disorder. In the other extreme, $r=0.5$ produces a fully disordered model but at $T=\infty$, which is again trivial. Furthermore, the relationship between disorder and temperature in this model puts it on top of the so-called Nishimori line \cite{nishimori2001statistical}, for any $r$. This means that the model passes from an uninformative phase (with $q\sim 0$) to a ferromagnetic phase (with $q>0$) without crossing any spin-glass phase.}. In particular, with the following transformation
\[\forall _{(i,j):s_i=-1} [(s_i=1) \text{ and }\forall_{j \in \partial i}(J_{ij}=-J_{ij})] \]
we redefine all nodes to be of type +1 while changing the $J_{ij}$ to have the same statistic as in Eq.\ref{eq:probJij}. Hence, the transformed ground truth network $G$ has $s_i=1$ for any $i$, and the fraction of misleading links observed by the individual is $r$. In this way, the average overlap in Eq.\ref{eq:overlap} is the sum of the opinion's expected values divided by $N$. In the physics literature, this is often called a \textit{gauge transformation} \cite{nishimori2001statistical}.

In the following sections, we start by studying the equilibrium solution (i.e., for $\tau \to \infty$) of the three heuristic rules (in the RN case, we obtain an analytical solution even for a finite thinking time). Finally, with numerical simulations, we compare their performances for any $\tau$.

\subsection{Equilibrium belief propagation process}
We first study analytically the properties of the BP process at the equilibrium, i.e., when $\tau\to \infty$.

\begin{figure}[h!]
    \centering
    \subfigure[]{\includegraphics[width=0.45\linewidth]{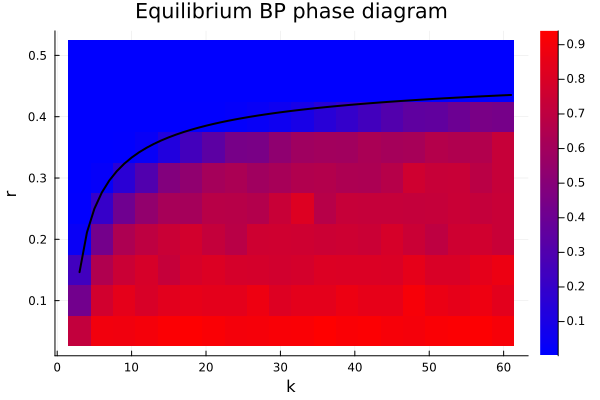}}
     \subfigure[]{\includegraphics[width=0.45\linewidth]{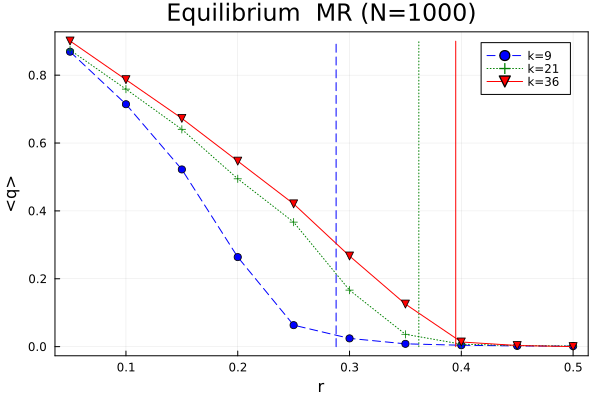}}
    \caption{(a) Phase diagram of the equilibrium BP process in the $[r,k]$ space (where $r$ is the noise and $k$ the average degree. Colors represent the values of the average overlap $\langle q \rangle$. Each pixel in the phase diagram is the average of 500 simulations of the BP process in a source network with $N=1000$ nodes. The solid line corresponds to Eq.\ref{eq:r_cBP}. (b) Average overlap $\langle q \rangle$ against noise $r$ in the equilibrium MR process for different average degree $k$. Each point is the average of 1000 simulations with $N=1000$ nodes. Vertical lines shows the theoretical values of the noise tipping point predicted by Eq.\ref{eq:MR_phase_transition} for the corresponding $k$.}\label{fig:BP_phase}
\end{figure}

Let us rewrite the message updating equation (\ref{eq:message}) in the following normalized form:
\begin{equation}\label{eq:message_updating}
    m_{i\to j}^+=\frac{1}{Z^*} \sum_{s=1,2}\phi^+(s)\prod_{z\in\partial i \setminus j}m_{z \to i}(s),
\end{equation}
where $\phi^+(s)$ is the $s$-th element of the first row of matrix $\phi_{J_{ij}}(i,j)$ in Eq.\ref{eq:phi_matrix} (we denote the second row as $\phi^-$), $m_{z\to i}(s)$ is the $s$-th element of the $m_{z\to i}$ vector, and $Z^*$ is the normalization constant given by
\begin{equation}
    Z^*:=\sum_{(a=\pm)}\sum_{s=1,2}\phi^a(s)\prod_{z\in\partial i \setminus j}m_{z \to i}(s).
\end{equation}
In this way, we have $m^-_{i\to j}=1-m^+_{i\to j}$. Now, let us denote the message in Eq.\ref{eq:message_updating} as a function of $\phi^+$ and all messages $m_{z\to i}$ for $z\in\partial i \setminus j$, i.e. $m^+_{i\to j}\equiv f(\phi^+, m_{z\to i}, z\in\partial i \setminus j)$. The distribution of $m_{i\to j}^+$ can be written recursively as
\begin{equation}\label{eq:m_distr}
 P(m_{i\to j}^+)=\mathbf{E}_{J^o}\Bigg[ \int \prod_{z\in\partial i \setminus j} dm^+_{z\to i} P(m_{z\to j}^+) \delta(m_{i\to j}^+-f)\Bigg],
\end{equation}
where the expected value is over the observed relation matrix $J^o$. At the equilibrium, for $N$ sufficiently large, messages for every neighbour of node $i$ are equal, i.e. $m^+_{z \to i} = m^+_{z' \to i}\equiv m^+$ for $z,z' \in \partial i$. Assuming that the number of neighbors of $i$ is exactly $k$, the expected value of $m^+_{i\to j}$ can be written as
\begin{equation}\label{eq:statio_BP}
   \int dm^+ P(m^+)m^+=
   \mathbf{E}_{J^o} \Bigg[\int \prod_{}^{k-1} dm^+ f(m^+,\phi^+)P(m^+)\Bigg].
\end{equation}
This equation is obtained by multiplying Eq.\ref{eq:m_distr} by $m^+$, integrating over all its possible values, and then applying the Dirac delta on the right-hand side. Since, for $t\to \infty$, the overlap is $q=2m^+-1$, Eq.\ref{eq:statio_BP} can be solved numerically to obtain the equilibrium overlap $q$ as a function of noise $r$. A trivial solution is when no information on nodes' signs can be retrieved from the messages interactions, i.e., when all messages are $1/2$. By performing a stability analysis of this trivial solution, we find that there exists a phase transition at a critical noise value $r_c$ below which the infinite-time overlap is non-zero. To show this, let us expand Eq.\ref{eq:statio_BP} around $m^+=1/2$. In particular, for some small $\epsilon$, we set $m^+=1/2+\epsilon$ and $m^-=1/2-\epsilon$. In this way, using $\int dm^+P(m^+)=1$ and writing explicitly the function $f$, Eq.\ref{eq:statio_BP} becomes
\begin{equation}
\frac{1}{2}+\epsilon=\frac{1}{Z^*} \mathbf{E}_{J^o}\Bigg[\frac{1}{2^{k-1}}+(k-1)[\phi^+(1)-\phi^+(2)]\epsilon\Bigg].
\end{equation}
Noting that the partition function is $Z^*=\frac{1}{2^{k-2}}$ (as $\phi^+(1)+\phi^+(2)=1$) and $\mathbf{E}_{J^o}[\phi^+(1)-\phi^+(2)]=(1-2r)^2$, the condition for which Eq.\ref{eq:statio_BP} has exactly one fixed point is
\begin{equation}\label{eq:r_cBP}
    (1-2r)^2<\frac{1}{k-1} \implies r_c=\frac{1}{2}-\frac{1}{2\sqrt{k-1}}.
\end{equation}
This stability analysis is equivalent to the replica symmetric computation of the transition temperatures in disorder systems \cite{dotsenko2005introduction}. When $r>r_c$, there is only one solution to Eq.\ref{eq:statio_BP} corresponding to the one with non-informative messages and hence with $q=0$. On the other hand, when $r<r_c$, the overlap is different from zero and, since we consider a positive seed node, we have $q>0$. Fig.\ref{fig:BP_phase}a shows the phase diagram of the equilibrium overlap in the BP process in the $[k,r]$ parameter space. It shows that numerical simulations are consistent with Eq.\ref{eq:r_cBP}.

This result implies that, when $N$ is sufficiently large, there exists a noise threshold above which an observer that lingers on the network for an infinite time is unable to extract any information about the sources' reliability. However, when $r<r_c$, the overlap limit is non-zero and the observer's opinions are not random. Note that, with the standard assumption that the BP algorithm determines the correct marginals probabilities \cite{decelle2011inference}, the above result is true for any algorithm: when the noise is larger than $r_c$, no heuristic rules or algorithms can detect the real nature of the information sources.  



\subsection{MR and RN processes} \label{MRBP}
As with the BP process, because of the strong non-linearity of the MR process, we treat analytically only its equilibrium state. With a stability analysis analogous to that performed for the BP heuristic, we again find that, for $N$ large enough, there is a noise threshold above which the individual's opinions for $\tau \to \infty$ are random (not correlated with the ground truth). This critical noise is given by (detailed calculations are shown in Appendix A):
\begin{equation}\label{eq:MR_phase_transition}
    r_c=\frac{1}{2}-\frac{2^{k-2}}{\sum_{i=\left \lfloor{k/2}\right \rfloor+1}^k \binom{k}{i}(2i-k)}.
\end{equation}

However, this result is valid only in the large $k$ limit. Whether there is a real phase transition for small average degrees is an open question. Nevertheless, even for small $k$, Eq.\ref{eq:MR_phase_transition} gives us a tipping point beyond which the majority rule is no longer efficient in extracting information from the source network, as shown in Fig.\ref{fig:BP_phase}b.

Given the simplicity of the opinion formation rule of the RN process, we can solve it analytically even for finite $\tau$ and $N$. Let us first show the results for $\tau=1$, i.e., $t=N-1$, already known in the literature. Through a master equation approach, it is possible to show that, given a network size $N$ and a noise $r$, the expected value of $q$ is given by \cite{medo2021fragility,meng2021whom}
\begin{equation}\label{eq:qRN}
\langle q(N,r)\rangle^{RN}_{\tau=1}= \frac{\Gamma(N+1-2r)}{N\Gamma(N)\Gamma(2-2r)},
\end{equation}
which, as expected, simplifies to $1$ for $r=0$, and to 0 for $r=1/2$. Moreover, it decreases rapidly with $r$, as widely discussed in \cite{medo2021fragility}. By expanding Eq.\ref{eq:qRN} for large $N$, we have
\begin{equation}\label{eq:largeNqRN}
\langle q(N,r)\rangle^{RN}_{\tau=1} \sim N^{-2r}, 
\end{equation}
showing that, for $r>0$ and $\tau=1$, the overlap goes to 0 in the $N\to \infty$ limit. Hence, an observer who navigates the source network for a short time ($\tau=1$) using the RN rule does not get an encouraging result: his/her opinions would tend to be totally random for large network sizes, regardless of how small is the noise (excluding the $r=0$ case).

\begin{figure}[h!]
    \centering
   \subfigure[]{\includegraphics[width=0.45\linewidth]{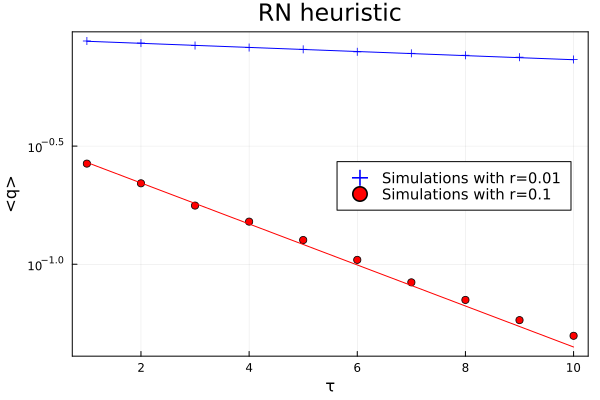}}
   \subfigure[]{\includegraphics[width=0.45\linewidth]{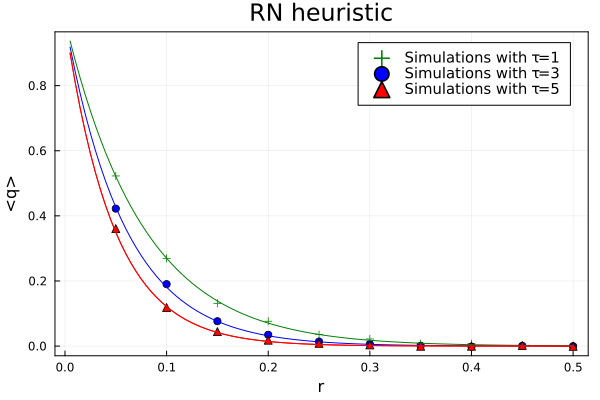}}
    \caption{(a) Expected value of the overlap $\langle q(N,r)\rangle^{RN}_{c>1}$ vs thinking time $\tau=t/(N-1)$ for the RN process with different level of noise $r$. (b) Expected value of the overlap $\langle q(N,r)\rangle^{RN}_{c>1}$ vs noise $r$ for different values of the thinking time $\tau$. In both panels: each point is the average of 1000 simulations with $N=1000$ nodes and average degree $k=200$; solid lines correspond to Eq.\ref{eq:RN}}\label{fig:RN}
\end{figure}

With a similar approach, we can also study the case $\tau>1$, showing that the observer's performance does not improve. In particular, we can write the expected value of $q$ for $\tau>1$ as (the detailed calculations are in appendix B):
\begin{equation}\label{eq:RN}
\langle q(N,r)\rangle^{RN}_{\tau>1}=q_1\left(1-\frac{2r}{N} \right)^{(\tau-1)(N-1)},  \end{equation}

where $q_1 := \langle q(N,r)\rangle^{RN}_{\tau=1}$. Fig.\ref{fig:RN} shows that Eq.\ref{eq:RN} is consistent with numerical simulations. As for the $\tau=1$ case, the overlap for $\tau>1$ decreases rapidly with noise. Moreover, it is straightforward to show that, when fixing $\tau$ and $r$, the behavior of Eq.\ref{eq:RN} with $N$ is the same as that of Eq.\ref{eq:largeNqRN}. At the same time, when $N$ and $r$ are fixed, Eq.\ref{eq:RN} decreases exponentially with the thinking time, asymptotically reaching 0 for $\tau\to \infty$. Notice also that all the above results on the RN process do not depend on the average degree $k$ (provided we neglect any finite-size effect). This was expected since, with the RN heuristic, the observer uses information from only one neighbor of the target node, regardless of its degree.

These findings show that the performance of the RN process not only decreases with the number of information sources but, unlike the BP and MR heuristics, also decreases with the time it spends for the observer to form opinions, regardless of how small the noise is. In short, the RN process, although it requires the least cognitive effort, is not a good strategy for forming opinions about a world of interconnected sources of information. Moreover, if the observer spends more time trying to infer the correct opinions with the RN rule, the performance is even worse.

\subsection{Performance comparison}
From the findings above, it is evident that, at the equilibrium, the RN process is the least efficient. Indeed, for $t\to \infty$, the overlap is 1/2 regardless of the noise level and the source network size. In the equilibrium BP and MR processes, however, $q$ can be larger than 1/2. In particular, in the large $N$ limit, there is a critical noise value below which $q>1/2$. Comparing the critical noise of the BP and the MR process, Eqs.\ref{eq:r_cBP} and \ref{eq:MR_phase_transition}, we observe that the former is always larger than the latter. It implies that when opinion formation with the MR strategy ceases to be informative, the BP strategy can still extract knowledge about the ground truth network. Said otherwise: the BP process is more tolerant to noise than the MR process.

\begin{figure}[h!]
    \centering
     \subfigure[]{\includegraphics[width=0.45\linewidth]{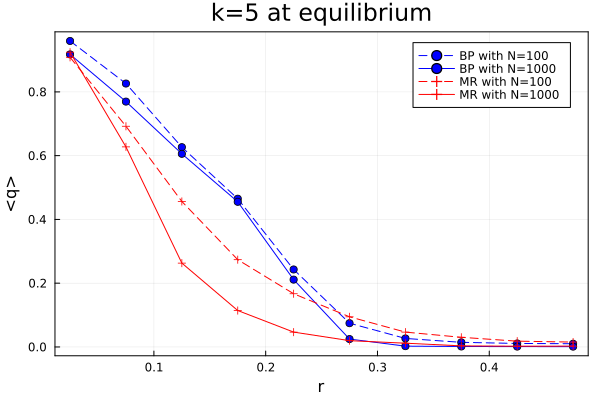}}
      \subfigure[]{\includegraphics[width=0.45\linewidth]{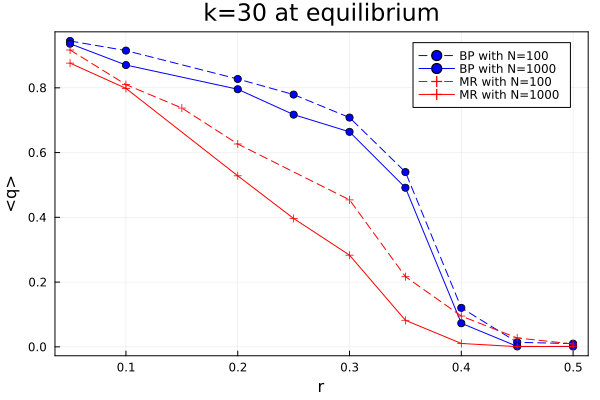}}
    \caption{(a) Equilibrium average overlap $\langle q \rangle$ against noise $r$ for both the MR and the BP process with different network sizes $N$ and with average degree $k=5$. (b) Same as panel (a) with average degree $k=30$. In both panels: each point is the average over 1000 realizations.}\label{fig:MRBPmag}
\end{figure}

Then, from numerical simulations, we find that the equilibrium overlap of the BP process is larger than that of the MP process even when $r<r_c$, for any $k$ and $N$ (see Fig.\ref{fig:MRBPmag}). These results are the consequence that in the BP heuristic the observer weighs the information from the neighbors of each node and harnesses the noise awareness to its advantage, as already mentioned. From these considerations, it is natural that the performance of the BP process at equilibrium is better than that of the MR process.

Furthermore, while it is obvious that both processes at equilibrium improve their performance as $k$ increases (as they can collect more information from the surroundings of each node), interestingly, the overlap for $t \to \infty$ of both processes (with $k$ fixed) decreases with $N$, until their convergence values. When the number of information sources is larger, the evaluation errors have more space to propagate and amplify the difference between the observer's opinions and the ground truth. This effect is most pronounced around the critical noise, where correlations between opinions in finite systems are most relevant. In light of these findings, we conjecture that, in random sources networks, any local heuristic rule makes individuals' opinions fragile to the number of information sources they must deal with.
However, when opinions are the result of the aggregation of information from multiple sources, the decrease in performance slows down for larger $N$ until it stops when the phase transition in $r_c$ can actually be observed. In fact, note that for finite $N$ the equilibrium overlap $q$ for both the BP and MR cases is a continuous and derivable function and that the calculations to obtain $r_c$ were performed in the so-called thermodynamic limit where the finite size effects can be neglected. The point about phase transitions being a property of infinite systems is a general clue from statistical physics \cite{kadanoff2009more, hartmann2006phase}.


\begin{figure}[h!]
    \centering
    \subfigure[]{\includegraphics[width=0.45\linewidth]{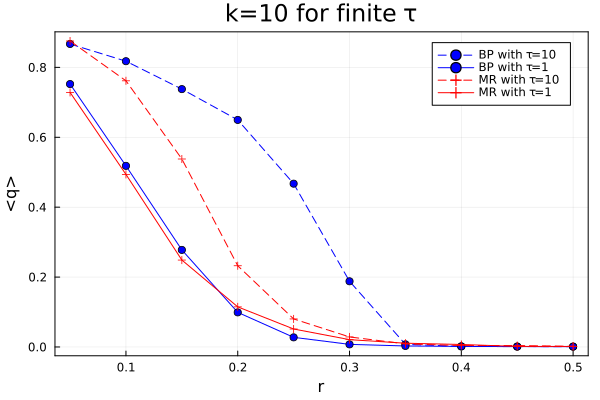}}
    \subfigure[]{\includegraphics[width=0.45\linewidth]{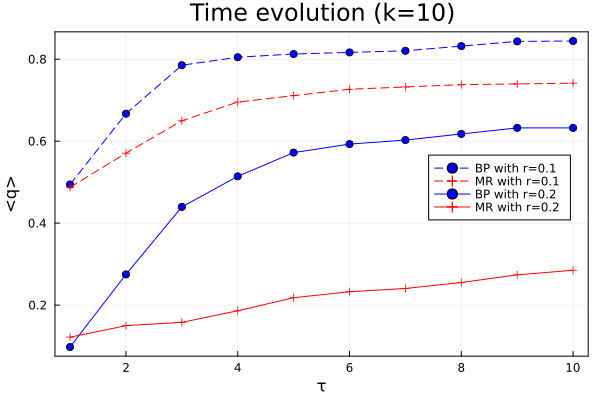}}
    \caption{(a) Equilibrium average overlap $\langle q \rangle$ against noise $r$ for both the MR and the BP process with different thinking times $\tau$. (b) Equilibrium average overlap $\langle q \rangle$ against thinking time $\tau$ for both the MR and the BP process with different noise $r$. In both panels: each point is the average over 1000 realizations with network size $N=1000$ and average degree $k=10$.}\label{fig:c1}
\end{figure}

Let us now study the more realistic case where the observer spends a limited time navigating the network of information sources. As shown in section \ref{MRBP}, the RN overlap is strongly sensitive to noise and decreases exponentially with $N$ up to the limiting value of 1/2 for $N\to \infty$. Fig.\ref{fig:c1}a compares the results for different $\tau$ in the BP and MR cases, showing that their overlaps decrease less dramatically with noise. Note that, for small $\tau$, the BP and MR processes have similar performances; instead, with large $\tau$, The BP strategy is significantly better than the MR one. This means that the benefit of spending time trying to form the correct opinions is larger in the case of the BP process. This is even more evident in Fig.\ref{fig:c1}b which shows the overlap as a function of $\tau$ for the BP and MR processes. 

Remember that, for the RN process, the performance decreases exponentially with time up to the limit value of 1/2 in its steady state, regardless of $r$. The reason for this decrease stems from the opinion formation rule that relies on a randomly chosen neighbor of the target node. Indeed, the noise in the sources network accumulates with the opinion updating, amplifying the difference between the observer's opinions and the ground truth.
Notably, this phenomenon is reversed for the BP and MR processes: as the thinking time increases, the overlap in both processes increases to its equilibrium value. This happens since both strategies aggregate information from all neighbors of a target node, with the effect of correcting errors instead of accumulating them. Again, from Fig.\ref{fig:c1}, we see that the BP performance improvement with the thinking time is much faster than that of the MR process. Indeed, in the BP process, the observer's opinions are probabilistic and allow uncertainties. As already mentioned, this results in a more careful aggregation of information which, as the thinking time increases, implies a higher capacity to identify reliable sources of information. In general, the more the opinion formation strategy is cognitively involved, the more it pays to think long and hard; conversely, lazier heuristic rules require more instinctive decisions. Finally, for the same reasons discussed in the previous section, the performance of both processes increases with $k$ and decreases with $N$, independently of $\tau$. Since we do not have an analytic solution away from the equilibrium, whether the BP and MR overlaps converge to a finite value when the network size becomes infinite remains an open question.


\section{Conclusion}

To summarize, we have reformulated the mechanism of the well-known belief propagation algorithm in the context of opinion formation. We proposed a probabilistic and local heuristic rule by which an individual extracts information from a system of interconnected sources to identify which among them are credible and which are not. Based on recent empirical evidence on the functioning of human reasoning, we claim that our model can realistically describe one of the mechanisms underlying the individuals' spontaneous opinion formation process. Thus, the next step would be to gauge it with real data and experiments. In addition, we have shown that our approach performs better than existing local heuristic rules in the literature: It is less sensitive to both structural noise and the size of the network of information sources; it also benefits more from the thinking time that individuals spend forming their opinions. However, we have shown that the efficiency of any local heuristic decreases as the number of information sources increases and that a phase transition is observed at equilibrium: even with our belief propagation approach, above a critical level of noise the individual's opinions are random.

These findings, although less pessimistic than the conclusions in \cite{medo2021fragility}, still hint that the reliability of the opinions of individuals with limited computation capabilities can be easily compromised, paving the way for the spread of misinformation. However, we provide a realistic framework that shows what are the conditions under which this is most likely. For this reason, we believe that our model is a useful tool for studying strategies to help people become properly informed.
At the same time, our work is strongly related and contributes to the literature on human learning with bounded rationality (see, for example, \cite{benjamin2019errors, galesic2021integrating}).

Our model can be extended in many directions.
First, note that we only addressed the case of a random source network. On the one hand, this has many advantages since, in random networks, it can be proven that the equilibrium belief propagation algorithm is a good approximation of the optimal Bayesian computation \cite{yedidia2003understanding}. However, in the real world we observe different types of networks, and therefore, in future research, it may be insightful to study how the performance of our belief propagation process varies in more complex network topologies. 

Second, for the sake of simplicity, we assumed that the noise, and thus the compatibility matrices, are the same for every link in the network. However, in the real world, different sources have different relationships, and thus the structural noise is also not the same. Hence, future models could investigate how different noise distributions among links in the source network affect the outcomes of opinion formation heuristic rules. Moreover, a limitation of our model over existing ones is that we must rely on the somehow strong assumption that the observer knows the true magnitude of the noise. However, Bayesian estimates of the human brain are imperfect, and this could be encoded in our model by defining another noise that affects the observer's estimate of the original noise. In this way, the compatibility matrices would be stochastic quantities. Another possibility to relax the assumption of perfect noise knowledge would be to design models by which the observer can ``learn'' and infer the noise in the network before the opinion formation process begins.

Finally, note that this paper investigates only one aspect of opinion formation, i.e., solitary opinion formation without social interaction. In contrast, previous literature has mostly focused on the social aspect of opinion propagation. Thus, the natural next step is to combine our heuristic rule with social influence.


\subsubsection*{Acknowledgements}
We thank M.\ Medo and Y.C.\ Zhang who helped to improve this paper by their comments.
\subsubsection*{Author contributions}
E.M.F. and A.L.C.  contributed equally to this work.

\subsubsection*{Competing Interests statement}
The authors declare no competing interests.

\appendix
\numberwithin{equation}{section}
\section{Derivation of the MR tipping noise}


Here, we derive Eq.\ref{eq:MR_phase_transition}. Let us denote by $P_i(t)$ the probability that, with the MR heuristic, the observer has the correct opinion on node $i$ at time $t$, i.e., $o_i(t)=(1,0)$ (under the gauge transformation). When the source network size $N$, the average degree $k$, and time $t$ are large enough, we can neglect correlations and assume $P_i(t)=P_j(t)\equiv P(t)$ for each $i$ and $j$. In this way, we can write

\begin{equation}\label{eq:MR_master}
    P(t+1)= P(t)\frac{N-1}{N}+ \frac{1}{N}\sum_{i=\left \lfloor{k/2}\right \rfloor+1}^k \binom{k}{i} W_t^i (1-W_t)^{k-i}.
\end{equation}

The first term on the right-hand side is the probability that the observer, at time $t$, has the correct opinion on the target node and that the latter is not chosen by the observer at time $t+1$ (recall that, for $t>N-1$, the observer chooses, at each time step, a random node). The second term is the probability that the target node is chosen at time $t+1$ and that the observer forms the correct opinion. Here, $W_t$ is the probability that the vote $v$ of a node toward the considered node is positive (at time $t$), i.e., $W_t \equiv (1-2r)P(t)+r$. Note that in Eq.\ref{eq:MR_master} we have implicitly made two further approximations: first, we have again assumed that the number of neighbors of each node exactly matches the average degree $k$; second, we have neglected the probability that $v=0$ (possible only when $k$ is even), i.e., we did not include in the sum the term $\frac{1}{2}\binom{k}{k/2}[W(1-W)]^{k/2}$ (however, in the critical noise calculation shown below, this term would not affect the result).

Now, at equilibrium we have $P(t+1)=P(t)\equiv P$ and $W_t\equiv W$. Thus, Eq.\ref{eq:MR_master} becomes
\begin{equation}\label{eq:stationary_MR}
P= \sum_{i=\left \lfloor{k/2}\right \rfloor+1}^k \binom{k}{i} W^i (1-W)^{k-i}. 
\end{equation}
We observe that when $r=1/2$ the only possible solution to this equation is $P=1/2$; on the other hand, when $r=0$, there are three possible solutions, i.e., $P=\pm 1$ and $P=1/2$. Therefore, we expect that there exists a critical noise $r=r_c$ such that, for $r<r_c$ and a positive seed node, $P>1/2$ and thus $q>0$ for $\tau \to \infty$. Indeed, expanding to first order Eq.\ref{eq:stationary_MR} around $P=1/2$, and solving the equation for $r$, we retreive Eq.\ref{eq:MR_phase_transition}.

\section{Derivation of the RN overlap}
Here, we derive Eq.\ref{eq:RN}. Let us define the probability $P(x,t)$ that, at time $t$, the number of correct opinions with the RN heuristic (and thus corresponding to $(1,0)$ due to the gauge transformation) is $x$. At each time step, $x$ can increase or decrease by one, or remain the same. Thus, the master equation of the RN process has the form
\begin{equation}\label{eq:masterRN}
P(x,t+1)=P(x-1,t)W(x-1\to x)
 +P(x+1,t)W(x+1\to x)+P(x,t)W(x\to x),
\end{equation}
where $W(x\to y)$ is the transition probability that the number of correct opinions becomes $y$ from $x$. To increase $x$ by one, the observer must pass over a node with an incorrect opinion and reevaluate it correctly. The probability that this occurs corresponds to $W(x-1 \to x)$, which, for a given $r$, is given by
\begin{equation}\label{eq:tran1}
 W(x-1 \to x)=\frac{N-x+1}{N}\left[\frac{x-1}{N-1}(1-r)+\frac{N-x}{N-1}r \right].
\end{equation}
Here, the first factor is the probability of choosing a node with an incorrect opinion. The second factor is the probability that the opinion on that node is reevaluated correctly. Similarly, we can write
\begin{equation}\label{eq:tran2}
 W(x+1 \to x)=\frac{x+1}{N}\left[\frac{x}{N-1}r+\frac{N-1-x}{N-1}(1-r) \right].
\end{equation}
The transition probability $W(x\to x)$, instead, is given by the probability that the observer does not change his/her past opinion on a node. This is
\begin{equation}\label{eq:tran3}
 W(x \to x)=\frac{x}{N}\left[\frac{x-1}{N-1}(1-r)+\frac{N-x}{N-1}r\right]\\
 + \frac{N-x}{N}\left[\frac{x}{N-1}r+\frac{N-1-x}{N-1}(1-r)\right].
\end{equation}
Now, by multiplying both sides of Eq.\ref{eq:masterRN} by $x$, summing over all possible $x$, and considering sufficiently large $N$, we can rearrange Eq.\ref{eq:masterRN} to obtain
\begin{equation}\label{eq:recursive}
\langle x(t+1)\rangle=\langle x(t)\rangle\left(1-\frac{2r}{N} \right)+r.
\end{equation}
Since the relation between $x$ and $q$ is $Nq=2x-1$, Eq.\ref{eq:recursive} is a recursive equation that must be solved with the initial condition
$$\langle x(N-1)\rangle=\frac{Nq_1-1}{2},$$
\\
where $q_1 := \langle q(N,r)\rangle^{RN}_{\tau=1}$.
Finally, from the solution of Eq.\ref{eq:recursive}, we can write the expected value of $q$ for $\tau>1$ as in Eq.\ref{eq:RN}.
Notice that this result better reproduces the numerical simulations when the information source network is not sparse (i.e. when $k$ is sufficiently large) so that the proportion of correct opinions in the transition probabilities of Eq.\ref{eq:tran1}, \ref{eq:tran2}, and \ref{eq:tran3} is better approximated.



\bibliographystyle{unsrt}
\bibliography{Manuscript.bib}

\end{document}